\documentclass[aps,prb,a4paper,10pt,twocolumn,showpacs,floatfix,superscriptaddress,amsmath,amsfonts,amssymb,preprintnumbers,longbibliography]{revtex4-1}
\usepackage{float}

\usepackage{dcolumn,graphicx,color,booktabs,microtype,afterpage}
\graphicspath{{./}{figure/}}
\usepackage[charter,greekuppercase=italicized]{mathdesign}
\usepackage{sidecap}
\renewcommand{\tablename}{Table}
\makeatletter\renewcommand{\fnum@figure}[1]{\figurename~\thefigure.~}\makeatother
\makeatletter\renewcommand{\fnum@table}[1]{\tablename~\thetable.}\makeatother

\newcount\hh \newcount\mm
\hh=\time \divide\hh by 60
\mm=\hh \multiply\mm by 60 \mm=-\mm
\advance\mm by \time
\def\now{\number\hh:\ifnum\mm<10{}0\fi\number\mm}

\usepackage[colorlinks,plainpages=false,linkcolor=blue,urlcolor=blue,citecolor=blue,pdfpagemode=UseNone,pdfstartview=FitBH]{hyperref}

\begin{document}

\makeatletter\renewcommand{\ps@plain}{%
\def\@evenhead{\hfill\itshape\rightmark}%
\def\@oddhead{\itshape\leftmark\hfill}%
\renewcommand{\@evenfoot}{\hfill\small{--~\thepage~--}\hfill}%
\renewcommand{\@oddfoot}{\hfill\small{--~\thepage~--}\hfill}%
}\makeatother\pagestyle{plain}

\preprint{\textit{Preprint: \today, \now.}} 

%
%
\title{Nodeless superconductivity in the noncentrosymmetric Mo$_3$Rh$_2$N superconductor: a $\mu$SR study}
\author{T.\ Shang}\email[Corresponding authors:\\]{tian.shang@psi.ch}
\affiliation{Laboratory for Multiscale Materials Experiments, Paul Scherrer Institut, Villigen CH-5232, Switzerland}
\affiliation{Swiss Light Source, Paul Scherrer Institut, Villigen CH-5232, Switzerland}
\affiliation{Institute of Condensed Matter Physics, \'Ecole Polytechnique F\'ed\'erale de Lausanne (EPFL), Lausanne CH-1015, Switzerland.}
\author{Wensen\ Wei}
\affiliation{Anhui Province Key Laboratory of Condensed Matter Physics at Extreme Conditions, High Magnetic Field Laboratory of the Chinese Academy of Sciences, Hefei 230026, People's Republic of China}
\author{C.\ Baines}
\affiliation{Laboratory for Muon-Spin Spectroscopy, Paul Scherrer Institut, CH-5232 Villigen PSI, Switzerland}
\author{J.\ L.\ Zhang}
\affiliation{Anhui Province Key Laboratory of Condensed Matter Physics at Extreme Conditions, High Magnetic Field Laboratory of the Chinese Academy of Sciences, Hefei 230026, People's Republic of China}
\author{ H.\ F.\ Du}
\affiliation{Anhui Province Key Laboratory of Condensed Matter Physics at Extreme Conditions, High Magnetic Field Laboratory of the Chinese Academy of Sciences, Hefei 230026, People's Republic of China}%
\author{M.\ Medarde}
\affiliation{Laboratory for Multiscale Materials Experiments, Paul Scherrer Institut, Villigen CH-5232, Switzerland}
\author{M.\ Shi}
\affiliation{Swiss Light Source, Paul Scherrer Institut, Villigen CH-5232, Switzerland}

\author{J.\ Mesot}
\affiliation{Paul Scherrer Institut, CH-5232 Villigen PSI, Switzerland}
\affiliation{Institute of Condensed Matter Physics, \'Ecole Polytechnique F\'ed\'erale de Lausanne (EPFL), Lausanne CH-1015, Switzerland.}
\affiliation{Laboratorium f\"ur Festk\"orperphysik, ETH Z\"urich, CH-8093 Zurich, Switzerland}
\author{T.\ Shiroka}
\affiliation{Laboratorium f\"ur Festk\"orperphysik, ETH Z\"urich, CH-8093 Zurich, Switzerland}
\affiliation{Paul Scherrer Institut, CH-5232 Villigen PSI, Switzerland}

\begin{abstract}
The noncentrosymmetric superconductor Mo$_3$Rh$_2$N, with $T_c = 4.6$\,K, adopts a $\beta$-Mn-type structure 
(space group $P$4$_1$32), similar to that of Mo$_3$Al$_2$C. Its bulk 
superconductivity was characterized by magnetization and heat-capacity measurements, while its microscopic 
electronic properties were investigated by means of muon-spin rotation and relaxation ($\mu$SR). 
The low-temperature superfluid density, measured via transverse-field (TF)-$\mu$SR, evidences 
a fully-gapped superconducting state with 
$\Delta_0 = 1.73\,k_\mathrm{B}T_c$, very close to 
1.76\,$k_\mathrm{B}T_c$ -- the BCS gap value for the weak coupling case, and a 
magnetic penetration depth $\lambda_0 = 586$\,nm. 
The absence of spontaneous magnetic fields below the onset of superconductivity, 
as determined by zero-field (ZF)-$\mu$SR measurements, hints at a preserved time-reversal 
symmetry in the superconducting state. Both TF-and ZF-$\mu$SR results evidence a spin-singlet pairing in Mo$_3$Rh$_2$N. 
\end{abstract}



\maketitle\enlargethispage{3pt}

\vspace{-5pt}
%
\emph{Introduction.} The current research interest in superconductivity (SC) involves either studies 
of high temperature superconductors (such as cuprates or iron pnictides), or investigations of 
unconventional superconducting states. Superconductors with centrosymmetric crystal structures 
are bound to have either pure spin-singlet or spin-triplet pairings.\cite{Anderson1984} 
On the other hand, due to the relaxed space-symmetry requirement, noncentrosymmetric 
superconductors (NCSCs) 
may exhibit unconventional pairing.\cite{Bauer2012,smidman2017} 
A lack of inversion symmetry leads to internal electric-field gradients and, hence, to antisymmetric spin-orbit  
coupling (ASOC), which lifts the spin degeneracy of the conduction-band electrons. As a consequence, the 
superconducting order can exhibit a mixture of spin-singlet and spin-triplet pairing.\cite{Bauer2012,smidman2017,Gorkov2001} 

Of the many NCSCs known to date, however, only a few exhibit a mixed singlet-triplet pairing. Li$_2$Pt$_3$B and Li$_2$Pd$_3$B are two notable examples, where the mixture of singlet and triplet states can be tuned by modifying the ASOC through a Pd-for-Pt substitution.\cite{yuan2006,nishiyama2007} Li$_2$Pd$_3$B behaves as a fully gapped $s$-wave superconductor, whereas the enhanced ASOC turns 
Li$_2$Pt$_3$B into a nodal superconductor, with typical features of spin-triplet pairing. 
Other NCSCs may exhibit unconventional properties besides mixed pairing. For instance, CePt$_3$Si,\cite{bonalde2005CePt3Si} CeIrSi$_3$,\cite{mukuda2008CeIrSi3} and K$_2$Cr$_3$As$_3$,\cite{K2Cr3As3Pen,K2Cr3As3MuSR} exhibit line nodes in the gap, while others such as LaNiC$_2$\cite{chen2013} and (La,Y)$_2$C$_3$,\cite{kuroiwa2008} show multiple nodeless superconducting gaps. In addition, due to the strong influence of ASOC, their upper critical 
fields can exceed the Pauli limit, as has been found in CePt$_3$Si\cite{bauer2004} and (Ta,Nb)Rh$_2$B$_2$.\cite{Carnicom2018}

Mo$_3$Al$_2$C forms a $\beta$-Mn-type crystal structure with space group 
$P$4$_1$32. Mu\-on\--spin rotation/relaxation ($\mu$SR), nuclear magnetic resonance (NMR), and specific heat studies have 
revealed that Mo$_3$Al$_2$C is a fully-gapped, strongly-coupled superconductor, which preserves time-reversal symmetry (TRS) 
in its superconducting state.\cite{bauer2010,Bauer2014} 
The recently synthesized Mo$_3$Rh$_2$N NCSC, a sister compound to Mo$_3$Al$_2$C, has been studied  
via transport and specific-heat measurements.\cite{Wei2016RhMoN} Yet, to date the microscopic nature of 
its SC state remains largely unexplored. DFT calculations suggest a strong hybridization between the Mo and Rh 4$d$-orbitals,  
reflecting the extended nature of the latter.\cite{Wei2016} The density of states (DOS) at the Fermi level $E_\mathrm{F}$, 
arising from the Rh and Mo 4$d$-orbitals, are comparable. This is in strong contrast with the Mo$_3$Al$_2$C case, where 
the DOS at $E_\mathrm{F}$ is mostly dominated by Mo 4$d$-orbitals.\cite{bauer2010,Karki2010} In the Mo$_3$Rh$_2$N case, 
the SOC is significantly enhanced by the replacement of a light element, such as Al, with one with a strong SOC, 
such as Rh. Considering that already 
Mo$_3$Al$_2$C exhibits unusual properties,\cite{bauer2010,Bauer2014} 
we expect the enhanced SOC to affect the superconducting properties of Mo$_3$Rh$_2$N, too. In Re$T$ ($T$= transition metal) alloys,\cite{Singh2014,Singh2017,Shang2018,Barker2018Re3Ta} whose DOS is dominated by the Re 5$d$-orbitals (with negligible 
contributions from the $T$ metal orbitals), even a robust increase in SOC --- from 
3$d$ Ti to 5$d$ Ta --- is shown to not significantly affect the superconducting properties. Conversely, similarly to the Li$_2$(Pd,Pt)$_3$B case, SOC effects are expected to be more important in Mo$_3$Rh$_2$N. Therefore, a comparative microscopic study of Mo$_3$Rh$_2$N 
vs.\ Mo$_3$Al$_2$C is very instructive for understanding the (A)SOC 
effects on the superconducting properties of NCSCs. Another goal of this study 
was the search for a possible TRS breaking in the superconducting state of Mo$_3$Rh$_2$N. 
 
In this paper, we report on the systematic magnetization, thermodynamic, and $\mu$SR investigation 
of the recently discovered Mo$_3$Rh$_2$N NCSC. In particular, zero- (ZF) and transverse-field (TF) $\mu$SR 
measurements allowed us to study the microscopic superconducting properties and to search for a possible TRS breaking below $T_{c}$ 
in Mo$_3$Rh$_2$N. 

\emph{Experimental details.} Polycrystalline Mo$_3$Rh$_2$N samples were synthesized by solid-state reaction 
and reductive nitridation methods, whose details are reported elsewhere.\cite{Wei2016RhMoN} The room-temperature x-ray powder diffraction
confirmed the $\beta$-Mn-type crystal structure, with no detectable extra phases.\cite{Wei2016RhMoN} The magnetization and heat capacity measurements were performed on a 7-T Quantum Design Magnetic Property Measurement System (MPMS) and a 9-T Physical Property Measurement System (PPMS). The bulk $\mu$SR measurements were carried out using the general-purpose surface-muon (GPS) and the low-temperature facility (LTF) instruments of the $\pi$M3 beamline at
the Swiss muon source of Paul Scherrer Institut, Villigen, Switzerland. For measurements on LTF, the samples were mounted on a silver plate using diluted GE varnish. The $\mu$SR data were analyzed by means of the \texttt{musrfit} software package.\cite{Suter2012}   
%
%

\begin{figure}[tb]
  \centering
  \includegraphics[width=0.48\textwidth]{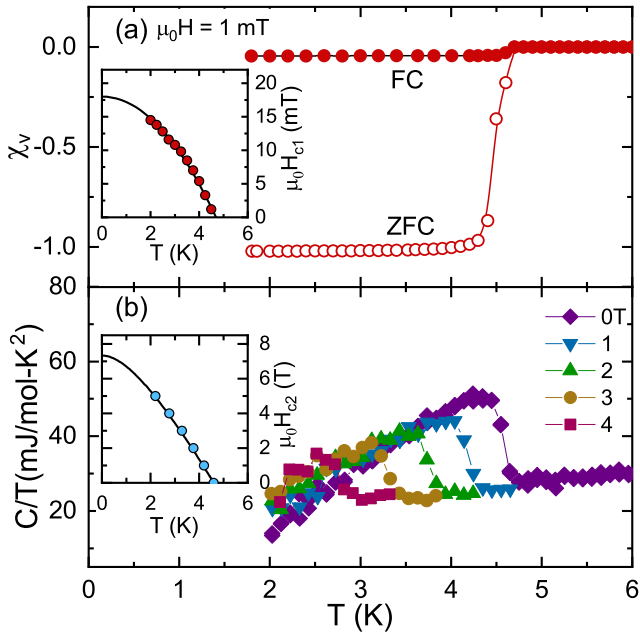}
  \caption{\label{fig:superconductivity}(a) Temperature dependence of magnetic 
  susceptibility $\chi(T)$ and (b) of specific heat $C(T)/T$ for Mo$_3$Rh$_2$N. 
  The inset in (a) shows the estimated $\mu_{0}H_{c1}$ vs.\ temperature up to $T_c$, 
  the solid-line being a fit to $\mu_{0}H_{c1}(T) =\mu_{0}H_{c1}(0)[1-(T/T_{c})^2]$. 
  For each temperature, $\mu_{0}H_{c1}$ was determined from the value where $M(H)$ 
  deviates from linearity. The inset in (b) shows $\mu_{0}H_{c2}(T)$, 
  as determined from heat-capacity measurements in various applied fields, with the 
  solid-line being a fit to the WHH model without spin-orbit scattering.}
\end{figure}
%
{\it Characterizing bulk superconductivity}. The magnetic susceptibility of Mo$_3$Rh$_2$N was measured using both field-cooled (FC) and 
zero-field-cooled (ZFC) protocols in an applied field of 1\,mT. As shown in Fig.~\ref{fig:superconductivity}(a), 
the ZFC-susceptibility indicates bulk superconductivity below $T_c = 4.6$\,K in Mo$_3$Rh$_2$N, consistent with the previously reported value.\cite{Wei2016RhMoN} The lower critical field $\mu_0$$H_{c1}$ was determined from the field-dependent magnetization $M(H)$, measured at various temperatures below $T_c$.  
The estimated $\mu_{0}$$H_{c1}(T)$ values are shown in the inset of Fig.~\ref{fig:superconductivity}(a). The solid-line represents a fit to $\mu_{0}H_{c1}(T) = \mu_{0}H_{c1}(0)[1-(T/T_{c})^2]$ and yields a lower critical field $\mu_{0}H_{c1}(0) = 18(1)$\,mT. The bulk superconductivity of Mo$_3$Rh$_2$N was further confirmed by heat capacity measurements [see Fig.~\ref{fig:superconductivity}(b)]. The specific heat, too, exhibits a sharp transition at $T_c$, which shifts towards lower temperature upon increasing the magnetic field. The sharp transitions ($\Delta T$ $\sim $ 0.3\,K) in both 
the specific-heat and magnetic-susceptibility data indicate a good sample quality. The derived $T_c$ values vs.\ the applied field are summarized in the inset of Fig.~\ref{fig:superconductivity}(b), from which the upper critical field $\mu_{0}H_{c2}$ was determined following the Werthamer-Helfand-Hohenberg (WHH) model.\cite{Werthamer1966} The solid-line in the inset of Fig.~\ref{fig:superconductivity}(b) represents a fit to the WHH model, without considering spin-orbital scattering, 
and gives $\mu_0$$H_{c2}(0)$ = 7.32(1)\,T, consistent with the previously reported value.\cite{Wei2016RhMoN}    


\begin{figure}[th]
	\centering
	\includegraphics[width=0.48\textwidth,angle=0]{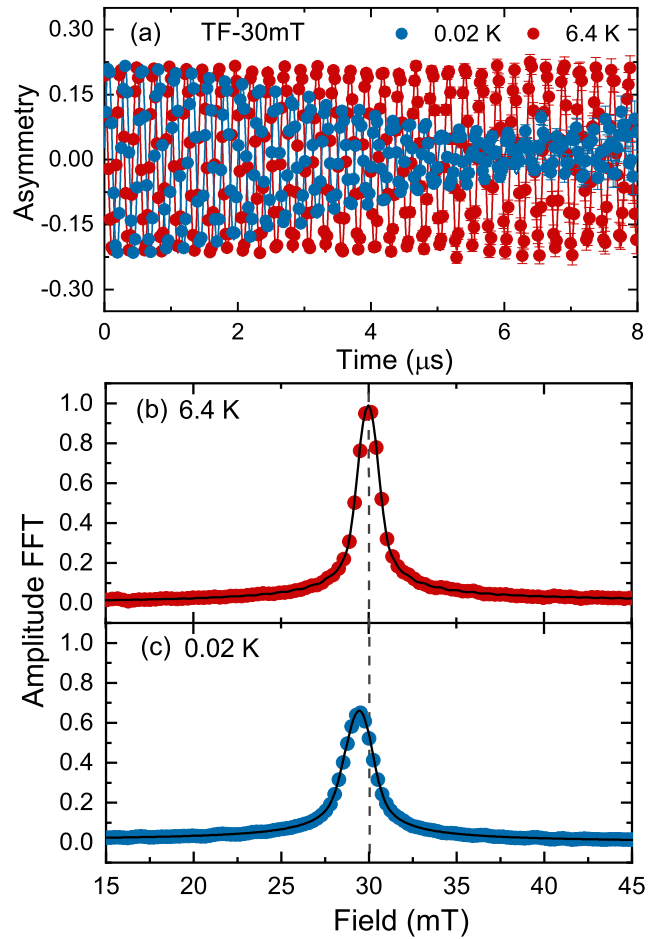}
	\vspace{-2ex}%
	\caption{\label{fig:TF_MuSR}(a) The Mo$_3$Rh$_2$N TF-$\mu$SR time 
	spectra, collected at 0.02\,K and 6.4\,K in an applied field of 
	30\,mT, show very different relaxation rates. Fourier transforms 
	of the above time spectra at 6.4\,K (b) and 0.02\,K (c). The solid lines 
	are fits to Eq.~(\ref{eq:TF_muSR}) using a single Gaussian relaxation; 
	the dashed lines indicate the applied magnetic field. Note the clear 
	diamagnetic shift below $T_{c}$ in (c).}
\end{figure}
%

\emph{Transverse-field $\mu$SR.} To explore the microscopic superconducting properties of Mo$_3$Rh$_2$N, TF-$\mu$SR measurements were performed down to 0.02\,K.   
In order to track the additional field-distribution broadening due to the flux-line-lattice (FLL) in the mixed superconducting state, a magnetic field of 30\,mT [i.e., larger than the lower critical field $\mu_{0}H_{c1}$(0)] was applied at temperatures above $T_c$. The TF-$\mu$SR time spectra were collected at various temperatures up to $T_c$, following a field-cooling protocol. Figure~\ref{fig:TF_MuSR}(a) shows two representative TF-$\mu$SR 
spectra collected above (6.4\,K) and below $T_c$ (0.02\,K) on GPS and LTF, respectively. The observed phase shift between the two datasets is due to instrumental effects. The faster, FLL-induced decay in the superconducting state is clearly seen in the second case. The time evolution of the $\mu$SR-asymmetry is modeled by: 
\begin{equation}
\label{eq:TF_muSR}
A_\mathrm{TF} = A_\mathrm{s} \cos(\gamma_{\mu} B_\mathrm{s} t + \phi) e^{- \sigma^2 t^2/2} +
A_\mathrm{bg} \cos(\gamma_{\mu} B_\mathrm{bg} t + \phi).
\end{equation}
Here $A_\mathrm{s}$ and $A_\mathrm{bg}$ represent the initial muon-spin 
asymmetries for muons implanted in the sample and sample holder, respectively, 
with the latter not undergoing any depolarization. The $A_\mathrm{s}$/$A_\mathrm{TF}$ 
ratios were determined from the long-time tail of TF-$\mu$SR spectra at base 
temperature [see Fig.~\ref{fig:TF_MuSR}(a)],\footnote{Due to the complete 
decay of the sample-related asymmetry beyond 7\,$\mu$s, the residual signal 
is due to the background only.} and fixed to 0.88 (GPS) and 0.90 (LTF) for all the temperatures.
$B_\mathrm{s}$ and $B_\mathrm{bg}$ are the local fields sensed by implanted muons in the sample and sample holder, $\gamma_{\mu} = 2\pi \times 135.53$\,MHz/T is the muon gyromagnetic ratio, 
$\phi$ is the shared initial phase, and $\sigma$ is a Gaussian relaxation rate. 
%
\begin{figure}[th]
	\centering
	\includegraphics[width=0.48\textwidth,angle=0]{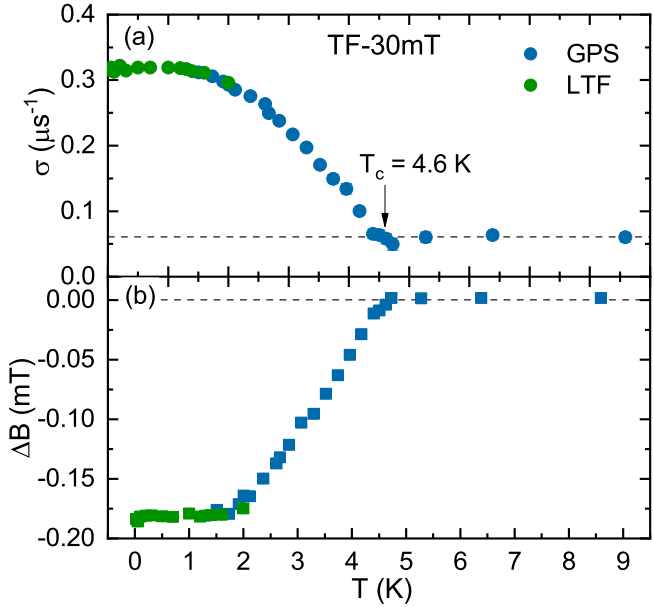}
	\vspace{-2ex}%
	\caption{\label{fig:sigma}Temperature dependence of (a) the muon-spin relaxation rate $\sigma(T)$ and (b) diamagnetic field shift $\Delta B(T)$ for Mo$_3$Rh$_2$N measured in an applied field of 30\,mT. Here $\Delta B = B_\mathrm{s} - B_\mathrm{bg}$, where $B_\mathrm{bg}$ is the same as the applied magnetic field.}
\end{figure}
%
The Gaussian nature of relaxation is clearly evinced from 
the fast-Fourier-transform (FFT) spectra shown in Fig.~\ref{fig:TF_MuSR}(b) and (c). In the mixed superconducting state, the faster decay of muon-spin polarization reflects the inhomogeneous field distribution due to the FLL, which causes the additional distribution broadening in the mixed state [see Fig.~\ref{fig:TF_MuSR}(c)]. In the superconducting state, the measured Gaussian relaxation rate includes contributions from both a temperature-independent relaxation due to nuclear moments ($\sigma_\mathrm{n}$) and the FLL ($\sigma_\mathrm{sc}$). The FLL-related relaxation 
can be extracted by subtracting the nuclear contribution according to $\sigma_\mathrm{sc}$ = $\sqrt{\sigma^{2} - \sigma^{2}_\mathrm{n}}$. The derived Gaussian relaxation rate and the diamagnetic field shift as a function of temperature are summarized in Fig.~\ref{fig:sigma}. The relaxation rate, shown in Fig.~\ref{fig:sigma}(a), is small and independent of temperature for $T > T_c$, but it starts to increase below $T_c$, indicating 
the onset of FLL and an increase in superfluid density. Concomitantly, a 
diamagnetic field shift appears below $T_c$ [see Fig.~\ref{fig:sigma}(b)].     

\begin{figure}[th]
	\centering
	\includegraphics[width=0.48\textwidth,angle=0]{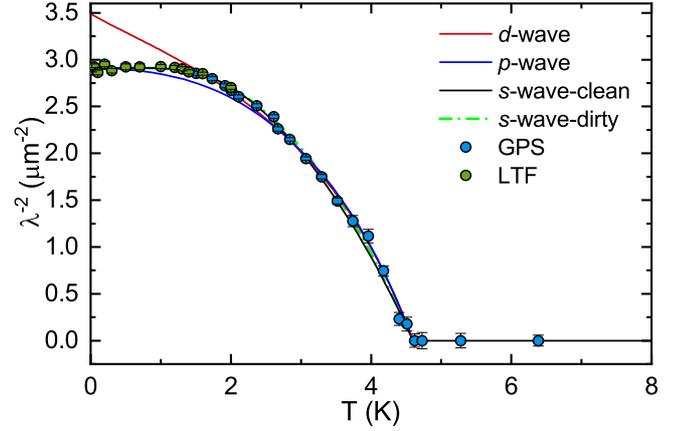}
	\vspace{-2ex}%
	\caption{\label{fig:lambda}  Superfluid density vs.\ temperature, as determined from TF-$\mu$SR measurements. 
		The different lines represent fits to various models, including $s$-, $d$-, and $p$-wave pairing (see text for details).}
\end{figure}

Since $\sigma_\mathrm{sc}$ is directly related to the magnetic penetration depth and the superfluid density ($\sigma_\mathrm{sc}$ $\propto$ $1/\lambda^2$), the superconducting gap value and its symmetry can be determined from the measured $\sigma_\mathrm{sc}$$(T)$. For small applied magnetic fields [$H_\mathrm{appl}$/$H_{c2}$ $\sim$ 0.004 $\ll$\,1], the magnetic penetration depth $\lambda$ can be calculated from:\cite{Barford1988,Brandt2003}
\begin{equation}
\label{eq:sig_to_lam}
\frac{\sigma_\mathrm{sc}^2(T)}{\gamma^2_{\mu}} = 0.00371\, \frac{\Phi_0^2}{\lambda^4(T)}.
\end{equation}
Figure~\ref{fig:lambda} shows the inverse square of the magnetic penetration depth (proportional to the superfluid density) 
as a function of temperature for Mo$_3$Rh$_2$N. To gain insight into 
the SC pairing symmetry in Mo$_3$Rh$_2$N, its temperature-dependent 
superfluid density $\rho_\mathrm{sc}(T)$ was further analyzed by using 
different models, generally described by:
%
\begin{equation}
\label{eq:rhos}
\rho_\mathrm{sc}(T) = 1 + 2\, \Bigg{\langle} \int^{\infty}_{\Delta_\mathrm{k}} \frac{E}{\sqrt{E^2-\Delta_\mathrm{k}^2}} \frac{\partial f}{\partial E} \mathrm{d}E \Bigg{\rangle}_\mathrm{FS},
\end{equation}
where $\Delta_\mathrm{k}$ is an angle-dependent gap function, $f = (1+e^{E/k_\mathrm{B}T})^{-1}$ is the Fermi function, and $\langle \rangle_\mathrm{FS}$ represents an average over the Fermi surface.\cite{tinkham1996} The gap function can be written as $\Delta_\mathrm{k}(T) = \Delta(T) g_\mathrm{k}$, where $\Delta$ is the maximum gap value and $g_\mathrm{k}$ is the angular dependence of the gap, equal to 1, $\cos2\psi$, and $\sin\theta$ for an $s$-, 
$d$-, and $p$-wave model, respectively. Here $\psi$ and $\theta$ are azimuthal angles.
%
%
The temperature dependence of the gap is assumed to follow 
$\Delta(T) = \Delta_0 \mathrm{tanh} \{ 1.82[1.018(T_\mathrm{c}/T-1)]^{0.51} \}$,\cite{tinkham1996} where $\Delta_0$, the gap value at zero temperature, is the only adjustable parameter. Note that the function $\Delta(T)$ is practically independent of the different models.

Three different models, including $s$-, $d$-, and $p$ waves, were used to 
describe the temperature-dependent superfluid density $\lambda^{-2}$$(T)$. 
By fixing the zero-temperature magnetic penetration depth 
$\lambda_\mathrm{0} = 586(3)$\,nm, the estimated gap values for the $s$- and $p$-wave 
model are 0.76(1) and 1.07(1)\,meV, respectively; while for the $d$-wave model, the estimated 
$\lambda_\mathrm{0}$ and gap value are 536(3)\,nm and 1.11(1)\,meV.
As can be seen in Fig.~\ref{fig:lambda}, the temperature dependence 
of the superfluid density is clearly consistent with a single fully-gapped 
$s$-wave model. 
In case of $d$- or $p$-wave models, a poor agreement with the measured 
$\lambda^{-2}$ values is found, especially at low temperature. The 
$s$-wave nature of SC is further confirmed by the temperature-independent behavior of $\lambda^{-2}(T)$ 
for $T < 1/3T_c$, which strongly suggests a nodeless superconductivity 
in Mo$_3$Rh$_2$N. Such conclusion is supported also by low-$T$ 
specific-heat data.\cite{Wei2016RhMoN}

Unlike the clean-limit case [see Eq.~(\ref{eq:rhos})], in the dirty limit 
the coherence length $\xi$ is much larger than the electronic mean-free 
path $l_\mathrm{e}$. In this case, in the BCS approximation, the temperature 
dependence of the superfluid density is given by:\cite{tinkham1996}
\begin{equation}
\label{eq:dirty}
\rho_\mathrm{sc}(T) = \frac{\Delta(T)}{\Delta_0} \mathrm{tanh} \left[\frac{\Delta(T)}{2k_\mathrm{B}T}\right].
\end{equation}
Following the above equation, the estimated gap value is 0.68(1)\,meV, 
slightly smaller than the clean-limit value, yet still in excellent agreement 
with the gap values extracted from low-$T$ specific-heat (0.67\,meV) and 
Andreev-reflection spectroscopy data (0.59\,meV).\cite{Wei2016RhMoN} 
Such `dirty'-nature of SC 
might reflect the large electrical resistivity ($\rho_0 = 0.48$\,m$\Omega$cm) 
and the small residual resistivity ratio (RRR $\sim$ 1) of Mo$_3$Rh$_2$N. 
The $2\Delta/\mathrm{k}_\mathrm{B}T_{c}$ ratios of about 3.46 (dirty limit) 
and 3.84 (clean limit) are both comparable to 3.53, the ideal value expected 
for a weakly-coupled BCS superconductor.
 
%
\begin{figure}[ht]
	\centering
	\includegraphics[width=0.48\textwidth,angle=0]{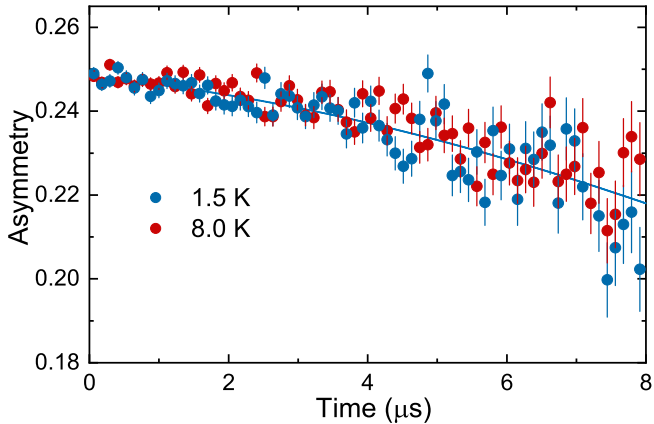}
	\vspace{-2ex}%
	\caption{\label{fig:ZF_muSR} Coinciding 
	ZF-$\mu$SR spectra in the 
	superconducting (1.5\,K) and the normal state (8\,K) show that in Mo$_3$Rh$_2$N 
	the TRS is preserved. Both spectra show only a weak muon-spin depolarization, 
	but no visible differences. The solid line is a fit to the 1.5-K spectra 
	by means of Eq.~(\ref{eq:KT_and_electr}), as described in the text.}
\end{figure}

\emph{Zero-field $\mu$SR.} We performed also ZF-$\mu$SR measurements, in order 
to search for a possible TRS breaking in the superconducting state of Mo$_3$Rh$_2$N. The large muon gyromagnetic ratio, combined with the availability of 100\% spin-polarized muon beams, make ZF-$\mu$SR a very sensitive probe for detecting small spontaneous magnetic fields. This technique has been successfully used to detect the TRS breaking in 
the superconducting states of different types of materials.\cite{Hillier2009,Singh2014,Barker2015,Luke1998,aoki2003,Shang2018}  Normally, in the absence of external fields, the onset of SC does not imply changes in the ZF muon-spin relaxation rate. However, if the TRS is broken, the onset of tiny spontaneous currents gives rise to associated (weak) magnetic fields, readily detected by ZF-$\mu$SR as an increase in muon-spin relaxation rate. 
Given the tiny size of such effects, we measured the ZF-$\mu$SR with high statistics in both  the normal and the superconducting phases. Representative ZF-$\mu$SR spectra collected above (8\,K) and below (1.5\,K) $T_c$ for Mo$_3$Rh$_2$N are shown in Fig.~\ref{fig:ZF_muSR}. For non-magnetic materials, in the absence of applied fields, the relaxation is mainly determined by the randomly oriented nuclear moments, which can be described by a Gaussian Kubo-Toyabe relaxation function $G_\mathrm{KT} = \left[\frac{1}{3} + \frac{2}{3}(1 -\sigma^{2}t^{2})\,\mathrm{e}^{(-\frac{\sigma^{2}t^{2}}{2})}\right] $.\cite{Kubo1967,Yaouanc2011} The ZF-$\mu$SR spectra of Mo$_3$Rh$_2$N can be modeled by adding a Lorentzian relaxation $\Lambda$ to the Kubo-Toyabe function:  
\begin{equation}
\label{eq:KT_and_electr}
A_\mathrm{ZF} = A_\mathrm{s} G_\mathrm{KT} \mathrm{e}^{-\Lambda t} + A_\mathrm{bg}.
\end{equation}
%
%
%
Here $A_\mathrm{s}$ and $A_\mathrm{bg}$ are the same as in the TF-$\mu$SR case 
[see Eq.~(\ref{eq:TF_muSR})].  
The resulting fit parameters are summarized in Table~\ref{tab:parameters}. The weak Gaussian and Lorentzian relaxation rates reflect the small value of Mo$_3$Rh$_2$N nuclear moments. The relaxations show very similar values in both the normal and the superconducting phase, 
as demonstrated by a lack of  
visible differences in the ZF-$\mu$SR spectra above and below $T_c$. 
\begin{table}[th]
	\centering
	\caption{\label{tab:parameters}
	Fit parameters extracted from ZF-$\mu$SR data for Mo$_3$Rh$_2$N 
	(collected above and below $T_c$) by using the Eq.~(\ref{eq:KT_and_electr}) 
	model.}
	%
	\begin{ruledtabular}	
		\begin{tabular}{lcc}
			Temperature               & 1.5\,K                      & 8\,K                    \\ \hline
	    	$A_\mathrm{s}$            & 0.24814(83)                 & 0.24833(73)             \\
	    	$\sigma$($\mu$s$^{-1}$)   & 0.0366(69)                  & 0.0379(58)              \\
	    	$\Lambda$($\mu$s$^{-1}$)  & 0.0069(32)                  & 0.0047(28)              \\
	    	$A_\mathrm{bg}$           & 0.01985(83)                 & 0.01987(73)             \\
	    \end{tabular}	
	\end{ruledtabular}
\end{table} 
%
%
%
This lack of evidence for an additional $\mu$SR relaxation below $T_c$, 
implies that TRS is preserved in the superconducting state of Mo$_3$Rh$_2$N. 
Since TRS is preserved also in the Mo$_3$Al$_2$C sister compound, this 
explains the many common features shared by these two $\beta$-Mn-type 
NCSCs.\cite{Bauer2014} 

\emph{Discussion.} Since the admixture of spin-singlet and spin-triplet pairing depends on the strength of ASOC,\cite{Gorkov2001} the latter plays an important role in determining the superconducting properties of NCSCs. An enhanced ASOC can turn a 
fully gapped $s$-wave superconductor into a nodal superconductor, with 
typical features of spin-triplet pairing, as exemplified by the Li$_2$(Pd,Pt)$_3$B case. However, a larger SOC is not necessarily the only requirement for a  
larger ASOC and an enhanced band splitting $E_\mathrm{ASOC}$, since the latter two depend also on the specific crystal- and electronic structures. All
4$d$-Rh, -Ru and 5$d$-Ir are heavy SOC metals, but their ASOC-related band splittings $E_\mathrm{ASOC}$ are relatively small in some materials. For example, the expected $E_\mathrm{ASOC}$ values for Ce(Rh,Ir)Si$_3$, LaRhSi$_3$, Rh$_2$Ga$_9$, and Ru$_7$B$_3$ are less than 20\,meV (i.e., ten times smaller than in CePt$_3$Si or Li$_2$Pt$_3$B).~\cite{smidman2017} Therefore, their pairing states remain 
in the spin-singlet channel and all of them behave as fully-gapped superconductors. In $\beta$-Mn-type materials, like Mo$_3$Rh$_2$N, the replacement of a light metal such as Al by the heavy Rh does indeed increase the SOC, yet the $E_\mathrm{ASOC}$ still remains weak. Hence, the superconducting pairing is of spin-singlet type, in good agreement  
with both TF- and ZF-$\mu$SR results.
Further band structure calculations, which explicitly take into account 
the SOC effects, are needed to clarify this behavior. 
 
\emph{Summary.} We perfomed comparative $\mu$SR experiments to study the superconducting properties of NCSC Mo$_3$Rh$_2$N. Bulk 
superconductivity with $T_c = 4.6$\,K was characterized by magnetization and heat 
capacity measurements. The temperature variation of the superfluid density reveals nodeless superconductivity in Mo$_3$Rh$_2$N, which is
well described by an isotropic $s$-wave model and is consistent with 
a spin-singlet pairing. 
The lack of spontaneous magnetic fields below $T_c$ indicates 
that time-reversal symmetry
is preserved in the superconducting state of Mo$_3$Rh$_2$N.

%

This work was supported by the Schwei\-ze\-rische Na\-ti\-o\-nal\-fonds zur F\"{o}r\-de\-rung
der Wis\-sen\-schaft\-lich\-en For\-schung, SNF (Grants 200021-169455 and 206021-139082) and the National Natural Science 
Foundation of China (Grant No.\ 11504378). 

\bibliography{RhMoN_bib}

\end{document}